\documentstyle[12pt]{article}
\begin{document}
\title{How to construct quantum measure in Regge calculus?}
\author{V. Khatsymovsky \\
 {\em Institute of Theoretical Physics} \\
 {\em Box 803} \\
 {\em S-751 08 Uppsala, Sweden\thanks{Permanent adress (after 15
November 1993): Budker Institute of Nuclear Physics, Novosibirsk
630090, Russia}} \\

 {\em E-mail address: khatsym@rhea.teorfys.uu.se\thanks{Permanent
E-mail address (after 15 November): khatsym@inp.nsk.su}}}
\date{\setlength{\unitlength}{\baselineskip}
\begin{picture}(0,0)(0,0)
\put(9,13){\makebox(0,0){UUITP-21/93}}
\put(8,12){\makebox(0,0)}
\end{picture}
}
\maketitle
\begin{abstract}
We propose the following way of constructing quantum measure in Regge
calculus: the full discrete Regge manifold is made continuous in some
direction by tending corresponding dimensions of simplices to zero,
then functional integral measure corresponding to the canonical
quantization (with continuous coordinate playing the role of time)
can be constructed. The full discrete measure is chosen so that it
would result in canonical quantization one whatever coordinate is
made continuous.

This strategy is followed in 3D case where full discrete measure is
determined in such the way practically uniquely (in fact, family of
similar measures is obtained). Averaging with the help of the
constructed measure gives finite expectation values for links.
\end{abstract}
\newpage
\section{Introduction}

Regge calculus formulation of general relativity (suggested by Regge
in 1961 \cite{Regge}; see recent review \cite{Will-rev}) is most
suitable one for quantizing it, because of countability of the set of
field variables. The role of the latter is played by linklengths. At
the same time, contrary to usual lattice, Regge manifold is itself a
particular case of Riemannian manifold and it's discretization step
is itself field variable. On the other hand, any Riemannian manifold
can be approximated (in the sense of measure) with arbitrary accuracy
by Regge manifolds. Thus, not only Regge manifold is a particular
case of Riemannian one, but smooth Riemannian manifold is a limiting
case of Regge one. In this sense we may speak about two a'priori
equal in rights formulations; but if, for example, calculating
expectation values of linklengths we shall find them to be zero, this
would mean that spacetime is smooth and Regge calculus description is
inadequate. If, however, expectation values of linklengths will be
nonzero, one can say that spacetime is discretized dynamically. These
expectation values will be of order of Plank length (the only
dimensional parameter at our disposal). Besides that, continuous
symmetries will be restored when summing not only over possible
linklengths, but also over all different schemes of pairing different
points into links. So at macroscopic level we shall have smooth
manifold, while effectively there will be the UV cut off at Plankian
scale (in the considered 3D model the linklength VEV's will be just
nonzero).

Well, one should quantize in the framework of Regge calculus, but how
to construct measure? The simplest way is to fix it by hand, but more
interesting is to obtain the measure from first principles. Such
principle in quantum mechanics is, of course, that of canonical
quantization. In relativistic theory it is that of equivalence of
different coordinates. So we act as described in abstract: the
measure of interest should reduce to the canonical one whatever
coordinate is made continuous.

In 3D model such the program can be fulfilled giving a family of
similar measures satisfying above conditions. It is interesting that
it is impossible to achieve maximal equivalence of different
directions: each measure is characterized by some field of directions
along which links should not fluctuate. It proves that namely these
links are allowed to have timelike interval. So time proves to be
nonquantizable which seems to be very natural.

Before going into technical details, we should emphasize, first, that
we study only local structure of spacetime and restrict ourselves by
Euclidean topology. As for global topology degrees of freedom, their
dynamics should coincide \cite{Wael} with that described by the
continuum theory \cite{Witten}. Second, our measure may not coincide
with those resulting from another approaches to 3D gravity, e.g. with
those based on 3j-symbols \cite{Will}, since issue principles are
different (although such the coincidence, if proven, would be an
interesting thing). Third, formal expression for measure leads to
divergences in physical region; therefore one can hope to give it a
sense by passing to integration over complex variables. Nontrivial is
that such region in complex plane exists for it should satisfy
several conditions simultaneously, as we shall see further. Finally,
main results and consequences of this paper, although in the not
quite general notations, are contained in the earlier unpublished
author's work \cite{Kha}.

The paper is organized as follows. In the next section the process of
reducing Regge action to continuous time Hamiltonian form is
considered. In section 3 canonical quantization measure  in different
forms is considered. In section 4 we move in backward direction and
step by step recover full discrete measure from canonical one.
Section 5 discusses link averages.

\section{Hamiltonian form}

An issue point is triad-connection formulation of Regge calculus,
according to which 3D Regge action

\begin{equation}\label{S-l}
S(l)=\sum{l\varphi (l)}
\end{equation}
is presented in the form \cite{Kha1}

\begin{equation}\label{S-l-Omega}
S(l,\Omega)=\sum{l\arcsin{{\vec{l}*R(\Omega) \over
2l}}},~~~\vec{l}*R\stackrel{\rm def}{=}l^aR^{bc}\epsilon_{abc}
\end{equation}
The (\ref{S-l}) is sum over links of linklength weighted angle
defects $ \varphi (l)$ which are functions of linklengths $ l$. In
(\ref{S-l-Omega}) these angle defects  are expressed as functions of
new independent variables $ \Omega$ - orthogonal (SO(3) in Euclidean
and SO(2,1) in Minkowsky case) matrices defined on triangles. Here
defined on links

\begin{equation}\label{1}
R(\Omega)=\prod{\Omega^{\pm 1}}
\end{equation}
are curvature matrices  - path-ordered products of $ \Omega$'s on
triangles sharing given link. It is thought of each tetrahedron that
there is local frame associated to it. The vector $ \vec{l}$
corresponding to each link is defined in the frame of some one of
tetrahedrons sharing this link.

We aim at constructing at least some formal expressions for measure.
Upon solving this problem we should give any strict sense to the
result obtained as a rule for obtaining expectation values of
different quantities in physical Minkowsky spacetime. Let us start
from Euclidean theory. It proves to be convenient to pass from the
very beginning to integration in path integral over imaginary link
vectors:

\begin{equation}\label{2}
\vec{l}\Longrightarrow -i\vec{l}.
\end{equation}
This trick, while leaving all three coordinates equivalent, allows
one to solve several problems at once. First, after such
transformation each coordinate is suitable for using it as formal
time in canonical analysis. Second, we obtain factor $ i$ multiplying
action in the path integral exponential,
$$\exp{(iS)},$$
which makes all the intermediate  integrals conditionally convergent.
Third, rotational group remains SO(3); therefore integrations over
connections are finite which is important in extending measure to
include integrations over all the connections.

Let $ A,~B,~C,~...$ denote vertices in 3D Regge manifold, then $
(AB),~(ABC),~(ABCD)$ will denote link, triangle and tetrahedron,
respectively, containing vertices in round brackets. These brackets
will denote unordered sequences of vertices. More detailed notation
for vector of link $ (AB)$ will be $ \vec{l}_{(AB)}$, for connection
on a triangle $ (ABC)$ - $ \Omega_{(ABC)}$. To point out the
tetrahedron $ (ABCD)$ in the frame of which the link vector is
defined we shall write $$\vec{l}_{(AB)(CD)}.$$

Note that for each $ (AB)$ there is only one $ (ABCD)$ for which $
\vec{l}_{(AB)(CD)}$ enters action.

The form (\ref{S-l-Omega}) can be easily reduced to the canonical
one. For such the reduction be nonsingular one should issue from some
special type of arrangement of vertices  when passing to the
continuous time limit. Namely, each tetrahedron should contain
infinitesimal link. Divide the manifold into 2D leaves of vertices in
such a way that each tetrahedron lay between two neighboring leaves.
Label leaves by parameter $ t$ playing the role of time such that
it's difference between the neighboring leaves $ dt\rightarrow 0$ in
the continuous time limit. In the leaf at a moment $ t$ denote
vertices $ i,~j,~k,~...$. Then for each vertex $ i$ there should
exist both it's image $ i^+$ and pre-image $ i^-$ in subsequent and
preceding leaves, respectively, such that there are links $ (ii^+)$
and $ (ii^-)$ of the length $ O(dt)$; also links $
(ik),~(i^+k^+),~(i^-k^-)$ exist or does not simultaneously. In other
words, space between any two neighboring leaves is divided into
infinitesimal prisms which we shall denote just as their triangle
bases $ (ikl)$, see Fig.1.

\begin{eqnarray}
\label{prism}
\begin{picture}(60,130)(50,10)
\put(-100,90){\rm Fig.1. Infinitesimal 3-prism}
\put (100,20){\line(0,1){80}}
\put (100,20){\line(1,1){120}}
\put (100,20){\line(2,1){160}}
\put (100,20){\line(3,1){120}}
\put (100,20){\line(1,0){160}}
\put (100,100){\line(3,1){120}}
\put (100,100){\line(1,0){160}}
\put (220,60){\line(0,1){80}}
\put (220,60){\line(1,-1){40}}
\put (220,140){\line(1,-1){40}}
\put (220,140){\line(1,-3){40}}
\put (260,20){\line(0,1){80}}
\put (260,95){$~l^{+}$}
\put (220,140){$~k^{+}$}
\put (87,95){$i^{+}$}
\put (92,15){$i$}
\put (220,60){$~k$}
\put (260,15){$~l$}
\end{picture}\nonumber
\end{eqnarray}

Behavior of quantities $ \vec{l},~\Omega$ at $ dt\rightarrow 0$
depends on the type of simplices on which these are defined. We shall
conventionally call the simplices completely contained in one of the
leaf spacelike ones; those which contain vertices of two
(neighboring) leaves will be called timelike if their measure
(length, area) tends to zero as $ dt\rightarrow 0$ and diagonal
otherwise. Connections on spacelike and diagonal triangles are
discrete analogs of continuum connections for the parallel vector
transport at a distance $ O(dt)$ and can be naturally assumed to be
infinitesimal. For the same reasons local frames should be attributed
to infinitesimal prisms $ (ikl)$ rather then to the separate
tetrahedrons. Thus on spacelike simplices we have the quantities

\begin{equation}\label{3}
\vec{l}_{(ik)l},~~~\Omega_{(ikl)}\stackrel{\rm
def}{=}1+f_{(ikl)}dt,~~~(f_{(ikl)}^{ab}=-f_{(ikl)}^{ba})
\end{equation}
($\vec{l}_{(ik)l}$ is vector of link $ (ik)$ defined in the frame of
prism $ (ikl)$) and similar ones on diagonal simplices,

\begin{eqnarray}
\vec{l}_{(ik^+)l}&\stackrel{\rm
def}{=}&\vec{l}_{(ik)l}+\vec{N}_{kil}dt,\nonumber\\
\Omega_{(ik^+l)}&\stackrel{\rm def}{=}&1+f_{(ik^+l)}dt,\\
\Omega_{(ik^+l^+)}&\stackrel{\rm def}{=}&1+f_{(ik^+l^+)}dt,\nonumber
\end{eqnarray}
whereas, on the contrary, timelike objects carry infinitesimal $
\vec{l}$ and finite $ \Omega$:

\begin{equation}
\vec{l}_{(ii^+)(kl)}\stackrel{\rm
def}{=}\vec{N}_{i(kl)}dt,~~~\Omega_{(ii^+k)}\stackrel{\rm
def}{=}\Omega_{ik}.
\end{equation}

The resulting set of variables $ \vec{l},~\Omega,~\vec{N},~f$ seems
to be rather large to be applied to canonical quantization for it
contains superfluous variables. Let us exclude $ \vec{N}_{ikl}$ and
one of connections $ \Omega_{ik},~\Omega_{ki}$ for each link $ (ik)$.
This is easy to do, because $ N_{ikl}^a$ enters resulting at $
dt\rightarrow 0$ Lagrangian linearly as the factor at the finite part
of curvature on the diagonal link,
$$\epsilon_{abc}[(\Omega_{ik}\Omega_{ki}^{\delta_1})^{\delta_2}]^{bc}.
$$

Here $ \delta_1,~\delta_2=\pm 1,~~~\Omega^{-1}\equiv \bar{\Omega}$.
Thus $ \Omega_{ik}$ and $\Omega_{ki}$ are, up to possible inversion,
the same and one of them will be denoted as  $ \Omega_{(ik)}$, the
variable on unordered link. The $ \vec{N}_{ikl}$ becomes unnecessary
now.

The remaining set $ \vec{l},~\Omega,~\vec{N},~f$ is still too large,
but $ f_{ABC}$ enters Lagrangian being summed over interior of
infinitesimal prism:

\begin{equation}\label{h}
h_{(ikl)}\stackrel{\rm def}{=}\pm f_{(ikl)}\pm f_{(ik^+l)}\pm
f_{(ik^+l^+)}.
\end{equation}
The $ h_{(ikl)}$ is true analog of connection $ \omega_0$ of
continuum theory.

Now the set $ \vec{l},~\Omega,~\vec{N},~h$ would be appropriate to
consider $ \vec{l}_{(ik)},~\Omega_{(ik)}$ as conjugate variables and
$ \vec{N}_i,~h_{(ikl)}$ as Lagrange multipliers. Note that we have
denoted $ \vec{N}_{i(kl)}$ simply as $ \vec{N}_i$, because for any $
i$ there is only one $ (ikl)$ for which $ \vec{N}_{i(kl)}$ enters
Lagrangian. The only trouble is that it enters not as Lagrange
multiplier, but in the form

\begin{equation}\label{arcsin}
N_i\arcsin{{\vec{N}_i*R_i(\Omega) \over 2N_i}}
\end{equation}
Equations of motion for $ \vec{N}_i$, however, have the same
solution, $ R_i=1$, as if we would simply delete here symbol
'arcsin'\footnote{This implies convention that another solution, $
R_i=-1$, corresponding to angle defect $ \varphi=\pi$, should be
ignored.}.

This gives the Lagrangian in the form

\begin{eqnarray}
L&=&L_{\dot{\Omega}}+L_h+L_N\label{Lagr}\\
L_{\dot{\Omega}}&=&{1 \over
2}\sum_{(ik)}^{}{\vec{l}_{(ik)}*(\bar{\Omega}_{(ik)}\dot{\Omega}_{(ik)
})}\\
L_h&=&{1 \over 2}\sum_{(ikl)}^{}{\left[\sum_{{\rm cycle\, perm}\,
ikl}^{}{\varepsilon_{(ik)l}\Omega_{(ik)}^{\delta_{(ik)l}}\vec{l}_{(ik)
}}\right]*h_{(ikl)}}\\&&\stackrel{\rm def}{=}\vec{\bf
C}(h),~~~~~~\delta\stackrel{\rm def}{=}{1+\varepsilon \over
2}\nonumber\\
L_N&=&{1 \over 2}\sum_{i}^{}{\vec{N}_i*R_i}\\
&&\stackrel{\rm def}{=}{\bf
R}(\vec{N}),~~~R_i=\Omega_{(ik_n)}^{\varepsilon_{ik_n}}...\Omega_{(ik_
1)}^{\varepsilon_{ik_1}},~\varepsilon_{ik_j}=-\varepsilon_{(ik_j)k_{j-
1}}=\varepsilon_{(ik_j)k_{j+1}}.\nonumber
\end{eqnarray}
The form of kinetic term $ L_{\dot{\Omega}}$ uniform for each link is
obtained by making, if necessary, change of variables $
\Omega\rightarrow\bar{\Omega}$ and/or $ \vec{l}\rightarrow\Omega^{\pm
1}\vec{l}$. Here $ \varepsilon_{(ik)l},~\varepsilon_{ik}=\pm 1$ are
sign functions defined by the way of attributing to each link the
frame where it is defined. Up to notations, this describes the system
suggested for 2+1 gravity by Waelbroeck \cite{Wael} from symmetry
considerations. Now we have seen how it arises from Regge calculus.

\section{Canonical quantization measure}

How to write down canonical quantization anzats for Lagrangian
(\ref{Lagr}) in the functional integral form? Note for that , there
exists an analog of  Poisson brackets for specific form of the
kinetic term $ L_{\dot{\Omega}}$,

\begin{equation}
\{f,g\}={\partial f \over \partial\vec{l}}\times{\partial g \over
\partial\vec{l}}\cdot\vec{l}-{\partial f \over
\partial\vec{l}}*\left(\bar{\Omega}{\partial g \over
\partial\Omega}\right)+{\partial g \over
\partial\vec{l}}*\left(\bar{\Omega}{\partial f \over
\partial\Omega}\right),
\end{equation}
(summation over links is implied) such that

\begin{equation}
{\partial f\over \partial t}=\{f,H\},~~~H=-L_h-L_N.
\end{equation}
The $ \vec{\bf C},{\bf R}$ act via these brackets as rotations and
translations, respectively. These are I class constraints and form
closed algebra w.r.t. these brackets. Therefore formal functional
integral measure for this system follows immediately\footnote{Of
course, the same can be obtained in the more rigorous way
introducing, as in \cite{Wael}, the variables $
P_{ab}=l^c\Omega_a^{~f}\epsilon_{cfb}$ and $ \Omega^{ab}$ for each
link, canonically conjugated in usual sense. There will be $
\delta$-functions in the functional integral taking into account II
class constraints to which $ P,~\Omega$ are subject, $
\delta^6(\bar{\Omega}\Omega-1)\delta^6(\bar{\Omega}P+\bar{P}\Omega)$.
The invariant measure $ D\vec{l}{\cal D}\Omega$ just arises on
integrating out these $ \delta$'s. The $ \{f,g\}$ turn out to be
Dirac brackets w.r.t. this system of II class constraints.}:

\begin{equation}
d\mu=\exp{\left(i\!\!\int{\!\!L_{\dot{\Omega}}dt}\right)}\delta
(\vec{\bf C})\delta({\bf R})D\vec{l}{\cal
D}\Omega,~~~D\vec{l}\stackrel{\rm
def}{=}\prod_{(ik)}^{}{d^3\vec{l}_{(ik)}},~~~{\cal
D}\Omega\stackrel{\rm def}{=}\prod_{(ik)}^{}{{\cal D}\Omega_{(ik)}},
\end{equation}
where Lebesgue $ d^3\vec{l}$ and Haar $ {\cal D}\Omega$ measures
arise as the only ones invariant w.r.t. the symmetries generated by $
\vec{\bf C},~{\bf R}$. Raise $ \vec{\bf C},~{\bf R}$ from $
\delta$-functions to exponent with the help of Lagrange multipliers $
\vec{N},~h$ and rewrite $ d\mu$ as

\begin{equation}\label{d-mu}
d\mu=\exp{\left(i\!\!\int{\!\!Ldt}\right)}D\vec{l}D\vec{N}{\cal
D}\Omega Dh.
\end{equation}

The only thing that we have not done needed to reduce the measure to
the form standard for systems subject to I class constraints is to
fix the gauge and divide it by the volume of symmetry group generated
by constraints. Because of compactness of SO(3) such the procedure
for rotational subgroup is superfluous. As for the volume of
translational subgroup, it is infinite, and in some cases separating
it out can help in making functional integral convergent. To break
translational symmetry let us impose partial gauge condition $
\vec{\bf f}=0$ where

\begin{equation}
\vec{\bf f}=\{\vec{l}_{(ik)}-\vec{a}_{(ik)}|(ik)\in F\}.
\end{equation}
Here $ \vec{a}_{(ik)}$ are constant vectors, $ F$ is a set of links
arranged in nonintersecting and non-self-intersecting broken lines
passing through all the vertices, see Fig.2.

\begin{eqnarray}
\begin{picture}(20,80)(50,10)
\put(-140,60){\rm Fig.2. Gauge fixing: links of the set $F$}
\put(-103,45){\rm (solid lines) have fixed vectors.}
\put(100,20){\line(6,1){60}}
\put(160,30){\line(2,1){40}}
\put(200,50){\line(4,1){40}}
\put(100,21){\line(6,1){60}}
\put(160,31){\line(2,1){40}}
\put(200,51){\line(4,1){40}}
\put(100,20){\line(1,3){10}}
\put(110,50){\line(5,-2){50}}
\put(200,50){\line(1,2){10}}
\put(210,70){\line(3,-1){30}}
\put(110,50){\line(5,2){50}}
\put(160,70){\line(1,0){50}}
\put(210,70){\line(2,1){20}}
\put(110,80){\line(1,0){30}}
\put(140,80){\line(6,1){60}}
\put(200,90){\line(1,0){30}}
\put(110,51){\line(5,2){50}}
\put(160,71){\line(1,0){50}}
\put(210,71){\line(2,1){20}}
\put(110,81){\line(1,0){30}}
\put(140,81){\line(6,1){60}}
\put(200,91){\line(1,0){30}}
\put(160,30){\line(5,4){50}}
\put(160,30){\line(0,1){40}}
\put(110,50){\line(1,1){30}}
\put(110,50){\line(0,1){30}}
\put(160,70){\line(2,1){40}}
\put(160,70){\line(-2,1){20}}
\put(210,70){\line(-1,2){10}}
\put(230,80){\line(1,-2){10}}
\put(230,80){\line(0,1){10}}
\put(230,80){\line(-3,1){30}}
\put(100,51){$i_1$}
\put(163,59){$i_2$}
\put(209,59){$i_3$}
\put(233,79){$i_4$}
\end{picture}\nonumber
\end{eqnarray}

By the standard rule of Faddeev-Popov anzats after dividing by
translational group volume we get, apart from $ \delta(\vec{\bf f})$,
the Jacobian factor $ \det{\|\{{\bf R},\vec{\bf f}\}\|}$, for which
we have

\begin{equation}\label{det}
\det{\|\{{\bf R},\vec{\bf f}\}\|}^{-1}=\!\!\int{\!\delta(\{{\bf
R}(\vec{N}),\vec{\bf f}\})D\vec{N}}.
\end{equation}
But along any broken line we have the sequence of gauge conditions $
\vec{f}_j=\vec{l}_{i_ji_{j+1}}-\vec{a}_{i_ji_{j+1}}$ (where $
\{i_j|j=...,1,2,3,...\}$ is sequence of vertices, see Fig.2) and
product of $ \delta$-functions

\begin{equation}
\delta(\{{\bf
R}(\vec{N}),\vec{f}_j\})=\delta^3(\varepsilon_{i_ji_{j+1}}\Gamma_{i_ji
_{j+1}}\vec{N}_{i_j}+\varepsilon_{i_{j+1}i_j}\Gamma_{i_{j+1}i_j}\vec{N
}_{i_{j+1}})
\end{equation}
where $ \Gamma_{ik}$ is some SO(3) matrix (product of $\Omega$'s)
required to express $ \vec{N}_i$ in the frame where $ \vec{l}_{(ik)}$
is defined. The $ \vec{N}_{i_j}$ are consecutively integrated out in
(\ref{det}) under appropriate boundary conditions along each broken
line giving unity for the Jacobian factor of interest. The $
\delta(\vec{f})$ in the Faddeev-Popov anzats is eliminated by
integrations over $ d^3\vec{l}_{(ik)}$, $ (ik)\in F$.

As a result, the whole effect of our gauge fixing reduces to omitting
integration in the expression for canonical measure $ d\mu$ over
links in 2D leaf constituting the set of broken lines $ F$ which is
natural to call {\it field} of links, in analogy with continuum
vector field. From the viewpoint of coordinate equivalence it is
natural to expect that instead integration over timelike links can be
omitted as well. Indeed, due to conservation in time the constraint $
{\bf R}$ may be imposed only as initial condition; therefore $
D\vec{N}$ integration giving $ \delta({\bf R})$ at finite times is
not required and can be omitted.

Thus, in addition to the form (\ref{d-mu}) of canonical measure we
can work also with any other form where integration over some field
of links, spacelike or timelike ones, is omitted.

\section{Recovering the full discrete measure}

Now we can restore full discrete measure by going back and inserting
into $ d\mu$ the variables earlier eliminated by use of eqs. of
motion and/or by neglecting infinitesimal contributions to action.
The principle of coordinate equivalence will make such the
reconstruction practically unambiguous modulo choice of the field of
links inherent in already continuous time formalism; but even if we
issue from the form (\ref{d-mu}) of canonical measure with
integrations over all the links available, at some step when diagonal
links are reintroduced, we need to exclude integrations over some
field of diagonal links in order to avoid infinities. Therefore the
resulting expression for measure will not include different
coordinates in completely equivalent way. In this sense it would be
more correctly to speak about maximal coordinate symmetry, or about
symmetry in choosing field of links.

First remind that contribution of timelike links enters exponential
of the canonical measure not as term (\ref{arcsin}) in the original
discrete action, but with symbol 'arcsin' omitted. By coordinate
equivalence we should write down contribution of all other links to
exponential when passing to the full discrete measure also without
'arcsin'. Eventually, such the possibility is connected with local
triviality of solution to classical eqs. of motion of 3D gravity.
This allows us to introduce integration over diagonal links $
D\vec{l}_{(ik^+)}$ and over additional connection for each link so
that instead of $ {\cal D}\Omega_{(ik)}$ we shall have $ {\cal
D}\Omega_{ik}{\cal D}\Omega_{ki}$. Simultaneously we add terms $ {1
\over 2}\vec{l}_{(ik^+)}*R_{(ik^+)}$ in the exponential. Integrating
over $ D\vec{l}_{(ik^+)}$ results in $ \delta$-function
$$\delta^3(R_{(ik^+)}-\bar{R}_{(ik^+)}).$$
Integration of this $ \delta$ over one of $ {\cal D}\Omega_{ik}$, $
{\cal D}\Omega_{ki}$ due to invariance of Haar measure is equivalent
to integration over $ {\cal D}R_{(ik^+)}$ since $ R_{(ik^+)}$ has the
form $ \Gamma_1\Omega^{\pm 1}\Gamma_2$ where $ \Omega$ is $
\Omega_{ik}$ or $ \Omega_{ki}$ and $ \Gamma_1,~\Gamma_2\in{\rm
SO}(3)$. Therefore this integration gives constant.

However, such $ \delta$-functions should arise when integrating over
each of links meeting at any given vertex $ A$. The product of $
\delta$-functions of curvatures on all these links
$$\prod_{B}^{}{\delta^3(R_{(AB)}-\bar{R}_{(AB)})}$$
is singular, because arguments of $ \delta$'s are related by Bianchi
identities \cite{Regge}. If we exclude integration over only one or
two links, the same trouble will occur with product of $ \delta$'s of
curvatures on all the links having common vertex with those one or
two. Repeating this consideration we conclude that integration should
be excluded for set of links ${\cal F}$ forming nonintersecting and
non-self-intersecting broken lines passing through each vertex of the
now full discrete Regge manifold.

The above considered ways of gauge fixing in canonical measure $
d\mu$ just correspond to different choices (w.r.t. the set $ {\cal
F}$) of continuous coordinate playing the role of time. Namely, if we
restore full discrete measure by described above way from $ d\mu$
with $ D\vec{N}$ omitted, the set $ {\cal F}$ turns out to be
timelike, whereas omitting in $ d\mu$ integrations over spacelike set
$ F$ eventually gives $ {\cal F}=F$. If, finally, the most
symmetrical form for canonical measure (\ref{d-mu}) is used with
integrations over all links available, in order to avoid
singularities we should omit integration over some field of diagonal
links.

The integration element $ d^3\vec{N}_i$ should be converted, upon
rescaling it by a pover of $ dt$, into $ d^3\vec{l}_{(ii^+)}$. This
is the only possible way of recovering finite element $
\vec{l}_{(ii^+)}$ while keeping equivalence of different links.

It remains only to recover finite connections on spacelike and
diagonal triangles. These are now infinitesimal and enter theory
being summed over groups of spacelike and diagonal triangles into $
h_{(ikl)}$, see (\ref{h}). Here we note that due to SO(3)-invariance
we can rotate the frame in each tetrahedron so that in each such
group of connections all but one connection would become trivial;
say, nontrivial is $ \Omega_{(ikl)}$. This is a kind of choosing the
gauge in already full discrete theory. Correspondingly, in the
continuous time in each sum for $ h_{(ikl)}$ we have only one term, $
f_{(ikl)}$. The $ d^3f_{(ikl)}$ is converted, upon rescaling it by a
power of $ dt$, into $ {\cal D}\Omega_{(ikl)}$. It is the only way of
recovering $ \Omega_{(ikl)}$ if one stick to the equivalence of
different triangles. By the sense of choosing the gauge the result of
integrating over $ {\cal D}\Omega_{(ikl)}$ will not depend on the
diagonal triangle connections, and including integrations over them,
$ {\cal D}\Omega_{(ikl^+)}{\cal D}\Omega_{(ik^+l^+)}$, into the
measure results simply to some constant normalization factor. It is
important here that in our method of giving sense to ill-defined
theory rotational group is compact.

Thus we can write out the resulting form of the full discrete measure

\begin{eqnarray}
d{\cal M}_{\cal F}&=&\exp{(i\tilde{S})}\prod_{(AB)\not\in{\cal
F}}^{}{d^3\vec{l}_{(AB)}}\prod_{(ABC)}^{}{{\cal
D}\Omega_{(ABC)}},\label{d-M}\\
&&\tilde{S}\stackrel{\rm def}{=}{1 \over
2}\sum_{(AB)}^{}{l^a_{(AB)}\epsilon_{abc}R^{bc}_{(AB)}},
\end{eqnarray}
where, remind, links of $ {\cal F}$ form a set of nonintersecting and
non-self-intersecting broken lines passing through all the vertices.
Instead of consideration leading from $ d\mu$ to $ d{\cal M}_{\cal
F}$ one might having got $ d\mu$ guess about the form of $ d{\cal
M}_{\cal F}$ and then prove that such $ d{\cal M}_{\cal F}$ indeed
ensure correct form of $ d\mu$ whatever coordinate is made
continuous.

\section{Discussion. Link averages.}

Let us try to find some quantum averages using the obtained measure $
d{\cal M}_{\cal F}$. For nondegenerate manifold the number of
triangles is not less than the number of links (this follows from
simple combinatorial consideration with taking into account that each
link is shared by no less than three triangles). So it is tempting to
try to change variables so that some $ \Omega$'s were replaced by
curvatures $ R$. Since curvatures on links meeting at a vertex are
related by the Bianchi identity, we can use curvatures as independent
variables only on links not belonging to some set $ {\cal
F}^{\,\prime}$. This set is constructed completely analogously to $
{\cal F}$ and has the same structure (for example, one can take $
{\cal F}^{\,\prime}={\cal F}$). Now consider $
R_{(AB)},~(AB)\not\in{\cal F}^{\,\prime}$ as independent variables
instead of the same number of appropriately chosen $ \Omega$'s. The
dependence of given $ R$ on given $ \Omega$  (if any) has the form $
R=\Gamma_1\Omega^{\pm 1}\Gamma_2,~\Gamma_1,\Gamma_2\in{\rm SO(3)}$,
and, by properties of invariant measure, $ {\cal D}R={\cal D}\Omega$.
Subsequently substituting $ R$'s instead of $ \Omega$'s we find
Jacobian of this change being unity.

Now integration over $ {\cal D}\Omega$ in $ d{\cal M}_{\cal F}$ is
substituted by that over $ {\cal D}R^{\prime}{\cal D}\Omega^{\prime}$
where 'prime' means connections of the reduced set and curvatures on
links not belonging to $ {\cal F}^{\prime}$. But the measure still
not seems to simplify considerably: apart from simple terms $
\vec{l}*R^{\prime}$ in the exponential there are also $
\vec{l}_{(AB)}*R_{(AB)}$ at $ (AB)\in{\cal F}^{\prime}$ in which $
R_{(AB)}$ is bulky product of some $ R^{\prime},~\Omega^{\prime}$.
Remind, however, that integration over links of set $ {\cal F}$ is
absent, therefore these should be fixed by hand. Making use of this
circumstance to avoid unnecessary complications let us put $ {\cal
F}^{\prime}={\cal F}$ and rescale links of this set to zero. This
corresponds to the most detailed possible description of quantum
Regge manifold when coordinate measuring the distance along the field
of links $ {\cal F}$ is becoming continuous one and we have at our
disposal sections of the manifold at arbitrary value of this
coordinate.

Now we see that at such the description the measure $ d{\cal M}_{\cal
F}$ factorizes into the product of independent ones,

\begin{equation}\label{fact}
d{\cal M}_l=\exp{\left( {i \over 2}\vec{l}_{(AB)}*R_{(AB)}
\right)}d^3\vec{l}_{(AB)}{\cal D}R_{(AB)},~~~(AB)\not\in{\cal F},
\end{equation}
referred to separate links. There is also integration over $ {\cal
D}\Omega^{\prime}$, but $ \Omega^{\prime}$'s enter only through $
R_{(AB)},~(AB)\in{\cal F}^{\prime}$, and upon setting $
\vec{l}_{(AB)}=0$ this integration is trivial.

Here one should point out the apparent contradiction between separate
quantization of links, as given by (\ref{fact}), and triangle
inequalities. The matter is that the triad-connection formulation
provides the closure of links into triangles only as a consequence of
eqs. of motion for the connections, i.e. on classical level. Since
the functional integral over $ \Omega$'s is nonGaussian one, such the
closure is violated virtually. This circumstance is specific for
triad-connection formulation of quantum Regge calculus different from
usual link lengths one.

To find Euclidean quantum average of invariant $ l^{2k}$ one should
apply the measure $ d{\cal M}_l$ to $ (-i\vec{l})^{2k}=(-l^2)^k$
(remind that thus far we have worked in the region obtained from
Euclidean one by substitution $ \vec{l}\rightarrow -i\vec{l}$). Due
to SO(3)-invariance of $ d{\cal M}_l$ this is sufficient for
averaging any link function $ g(\vec{l})$. Integral over $
d^3\vec{l}$ gives derivatives of $ \delta$-function
$$\delta^3\left( {R-\bar{R} \over 2} \right)=\delta^3\left(
{\sin{\phi} \over \phi}\vec{\phi} \right)$$
where $ R\stackrel{\rm def}{=}\exp{(\epsilon^{abc}\phi_c)}$. Then $
\delta$-function is integrated out over
$${\cal D}R={\sin^2(\phi /2) \over 4\pi^2\phi^2}d^3\vec{\phi}$$
in the vicinity of $ \vec{\phi}=0$ (another root at $ \phi=\pi$
should be omitted according to our convention, see footnote 1 on page
5). The (normalized by $ <1>=1$) result gives

\begin{equation}\label{<l>}
<l^{2k}>_E=(-1)^k<l^{2k}>={4^{-k}\Gamma (2k+2)^2 \over \Gamma
(k+2)\Gamma (k+1)}
\end{equation}
for Euclidean quantum average. It is finite at $ k>-1$ and nonzero.
This allows to recover quantum average for any function of link:

\begin{eqnarray}
<g(\vec{l})>_E&\!\!\!\!=&\!\!\!\!\!\int{{do_l \over
4\pi}\int_{0}^{\infty}{g(\vec{l})\nu (l)dl}},\nonumber\\
&&\nu (l)={2l \over \pi}\int_{0}^{\pi}{\exp{\left( -{l \over
\sin{\varphi}} \right)d\varphi}}\label{<g>}
\end{eqnarray}
Thus, quantum average $ <\cdot>$ in our analytically continued by $
\vec{l}\rightarrow-i\vec{l}$ region where expressions for measure are
well-defined proves to be nonpositive. Remarkable, however, is that
on returning back to Euclidean region we get positively defined
measure.

Up to now, the word 'time' denoted simply continuous coordinate or
that which should be made such. If one continues formulas
(\ref{<l>}), (\ref{<g>}) to Minkowsky region, $ <l^2>$ remains
positive, although now $ l^2$ is indefinite quadratic form of
components $ l^a$ of signature (+,+,-). Let us now use the word
'time' in it's strict sense implying that timelike interval is
negative. In this sense fluctuating links are always spacelike ones.
Therefore timelike links may be only nonfluctuating ones, i.e. from
set $ {\cal F}$. Thus, time does not fluctuate, which is quite
natural. It is amusing that this fact arises here as consequence of
quantum Regge calculus. In addition, time cannot be more than
1-dimensional, since set $ {\cal F}$ pick out only one direction at
each point. In turn, occurrence of $ {\cal F}$ and it's
1-dimensionality can be traced to Bianchi identities. Thus,
1-dimensionality of time and it's nonquantizability in 3D spacetime
follow from the Bianchi identities!

To conclude, we see that since we get finite nonzero value for
linklength quantum averages for arbitrary scheme of pairing different
points into links, the Nature, indeed, prefer Regge calculus (at
least, in three dimensions, in Einstein type gravity, etc.). Opposite
case of zero averages would mean that simplices collapse giving
continuum theory.

\bigskip
I am grateful to prof. A. Niemi, S. Yngve and personnel of Institute
of Theoretical Physics at Uppsala University for warm hospitality and
support during the work on this paper.

\end{document}